\documentclass[review]{elsarticle}

\usepackage{lineno,hyperref}
\modulolinenumbers[5]

\journal{ArXiv}

\usepackage{epsfig,epstopdf}
\usepackage{subcaption}
\usepackage{float}
\usepackage{xcolor}

\usepackage{multirow}
\usepackage{multicol}
\usepackage{ragged2e}
\usepackage{subcaption}
\usepackage{float}

\usepackage{amsmath}
\usepackage{amssymb}
\usepackage{graphicx}

\DeclareMathOperator{\sgn}{sign}

\def\RE{\mathbb R}
\def\T{\mathcal T}
\def\G{\Gamma}

\begin{document}
\newtheorem{theorem}{Theorem}

\begin{frontmatter}

\title{Convergence time estimate and tuning of twisting algorithm}

%% Group authors per affiliation:
\author[ad1]{Ram\'on I. Verd\'es Kairuz}
\ead{rverdes@citedi.mx}
\author[ad2]{Yury Orlov}
\ead{yorlov@cicese.mx}
\author[ad1]{Luis T. Aguilar}
\ead{laguilarb@ipn.mx}
\address[ad1]{Instituto Polit\'ecnico Nacional---CITEDI, Avenida Instituto Polit\'ecnico Nacional 1310 Colonia Nueva Tijuana 22435 Tijuana, M\'exico}
\address[ad2]{Department of Electronics and Telecomunications, Mexican Scientific and Advanced Studies Center of Ensenada, Carretera Ensenada-Tijuana No. 3918,
	Zona Playitas, CP. 22860, Ensenada, M\'exico}

\begin{abstract}
Gain tuning is given for the twisting controller to ensure  that the closed-loop trajectories of the perturbed double integrator, initialized within a bounded domain and affected by uniformly bounded disturbances, settle at the origin in prescribed time.
\end{abstract}

%\begin{keyword}
%finite time stability \sep variable structure control \sep settling time estimate \sep twisting controller
%\end{keyword}

\end{frontmatter}

%\linenumbers

\setlength{\abovedisplayskip}{13pt}
\setlength{\belowdisplayskip}{13pt}
\setlength{\abovedisplayshortskip}{13pt}
\setlength{\belowdisplayshortskip}{13pt}
%%%%%%%%%%%%%%%%%%%%%%%%%%%%%%%%%%%%%%%%%%%%%%%%%%%%55555%%%%%%%%%%%%%%%%%%

\section{Introduction}

In   \cite{Orlov2012}, a settling time estimate of the perturbed double integrator
\begin{equation}
	\label{eq:SISO}
	\begin{split}
		\dot x_1&=x_2,\\
		\dot x_2&=u(x_1,x_2)+\omega(t),
	\end{split}
\end{equation}
driven by the twisting algorithm
\begin{eqnarray}
	\label{eq:u2def}
	u(x_1,x_2)&=&-\mu_2\sgn(x_1)-\mu_1\sgn(x_2),
\end{eqnarray}
was obtained in terms of the positive constant gains $\mu_1,\mu_2$, of the upper magnitude bound $N>0$ on the external disturbance $\omega(\cdot)$ such that
\begin{equation}
\left|\omega(t) \right|\leq N\quad \mbox{for all}\ t, \label{disturbound}
\end{equation}
and of the size $R>0$ of the initial domain
\begin{eqnarray}
	 \G_R&=&\left\lbrace x:V(x_1,x_2)\leq R\right\rbrace,
	\label{eq:G1def}
\end{eqnarray}
where $V(x_1,x_2)$ is the  positive definite function
\begin{equation}
	\label{eq:LyapTwisting}
	V(x_1,x_2)=\mu_2\left| x_1\right|+\frac{1}{2}x_2^2.
\end{equation}
The aim of this note is to further simplify the afore-mentioned estimate to suit it to control applications of the twisting algorithm in the closed-loop form. For convenience of the reader, the nomenclature of \cite{Orlov2012} is used throughout.

\section{Settling time estimate}

Given the parameters
	\begin{equation}
		\label{eq:Rbeta}
		R>0, \quad \beta>1,\quad
	%\end{equation}
%	let us set
%	\begin{equation}
%		\label{eq:EtaDelta}
%	%r_1=\sqrt{2R}, \quad	
%\eta\in (0,\frac{1}{\beta}] , \quad
\rho \in (0,1), \quad \delta> \frac{\sqrt{2R}(\beta +1)}{\beta -1}
	\end{equation}
	let us choose
	\begin{align}
		\label{eq:C1}
			\mu_1&> \frac{2\delta}{\T_s\sqrt{1-\beta^{-2}}}+N ,\\
			\label{eq:C2}
			\mu_2&> \max \left\lbrace \sqrt{\frac{R}{2}},\ \rho\sqrt{\frac{R}{2(1-\rho)}},\ \rho,\ \beta \mu_1,\ %\frac{\mu_1-N}{\eta},\
\mu_1+N\right\rbrace ,
	\end{align}
where $\T_s>0$ is a prescribed convergence time.

Setting the initial domain boundary
\begin{eqnarray}
	\partial \G_R&=&\left\lbrace x:V(x_1,x_2)= R\right\rbrace,
	\label{eq:BG1def}
\end{eqnarray}
 the following result is in order.
\begin{theorem}\label{teo}
Consider the perturbed double integrator \eqref{eq:SISO}, driven by the twisting algorithm \eqref{eq:u2def} and affected by an external disturbance \eqref{disturbound}. Let  $x(\cdot)=(x_1(\cdot),x_2(\cdot))^T \in \RE^2$ be an arbitrary solution of the closed-loop system \eqref{eq:SISO}--\eqref{eq:u2def}, initialized on the domain boundary
\eqref{eq:BG1def} of the size $R>0,$ and let the gains $\mu_1, \mu_2$ be chosen according to \eqref{eq:C1}--\eqref{eq:C2}, specified with $N,\T_s>0$ and the parameters $R,\beta,\rho,\delta$, satisfying \eqref{eq:Rbeta}. Then   $x(t)\equiv 0$ for all  $t\geq\T_s$, regardless of whichever external disturbance \eqref{disturbound} affects the closed-loop system.
\end{theorem}
\section{Proof of Theorem 1}
The proof follows the line of reasoning used in \cite{Orlov2012}. All relations, which are subsequently invoked from \cite{Orlov2012}, are  accordingly referenced.

\begin{figure}[H]
	\begin{center}
		\includegraphics[width=2.5in]{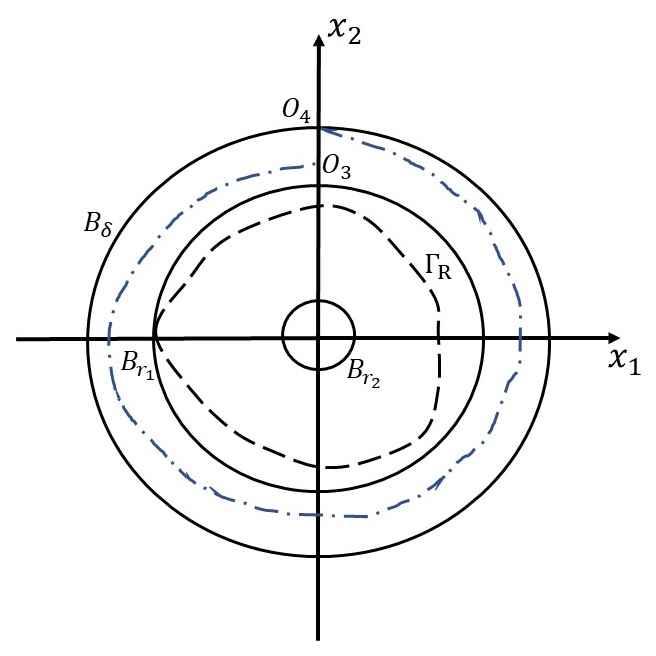}
	\end{center}
	\caption{Sets $\G_R$, $B_{r_1}$, $B_{r_2}$ and $B_{\delta}$.}
	\label{f:Balls}
\end{figure}

For later use, let us introduce embedded balls  $$B_{r_1}=\{x_1^2+x_2^2\leq  r_1^2\}\ \ \mbox{and}\  \  B_{r_2}=\{x_1^2+x_2^2\leq r_2^2\}\ \ \mbox{such that}\ \  B_{r_2}\subset \G_R\subset B_{r_1}.$$ The balls, thus introduced, are depicted in  Figure \ref{f:Balls}, reproduced from \cite{Orlov2012}.
The radiuses $r_1, r_2$ of such balls are established in \cite[relations (4.18) and (4.25)]{Orlov2012} to be
\begin{align}
	r_1&=\max\left\lbrace \frac{R}{\mu_2},\sqrt{2R}\right\rbrace\; \label{eq:r1}\\
	r_2&=\min\left\lbrace \frac{\rho R}{\mu_2},\sqrt{2R(1-\rho)}\right\rbrace\; \label{eq:r2}
\end{align}
with $\rho\in (0,1).$ Since  $\mu_2> \sqrt{\frac{R}{2}}$ due to \eqref{eq:C2}, relation \eqref{eq:r1} is simplified  to \cite[relation (6.6)]{Orlov2012}
\begin{equation}
	\label{eq:r1f}
	r_1=\sqrt{2R} \; 
\end{equation}
In addition, $\mu_2> \rho\sqrt{\frac{R}{2(1-\rho)}}$ by its choice \eqref{eq:C2}, and \eqref{eq:r2} is therefore simplified to \cite[relation (4.28)]{Orlov2012}
\begin{equation}
	\label{eq:r2f}
	r_2=\frac{\rho R}{\mu_2}.
\end{equation}
Taking into account that \eqref{eq:C2} results in $\mu_2>\rho$, it follows  that $r_2<R$.

 The convergence time $T_2$ of the closed-loop system \eqref{eq:SISO}--\eqref{eq:u2def}, initialized on the domain boundary \eqref{eq:BG1def}, is upper estimated by that of initialized at the intersection point $O_4$ of the vertical axis and the circle $\partial B_\delta=\{x_1^2+x_2^2=\delta^2\}$ of a (properly chosen) radius $\delta$, satisfying  \eqref{eq:Rbeta}; see Figure \ref{f:Balls}. As shown in \cite[p.480]{Orlov2012}, such a value of $\delta$  ensures that the resulting convergence time estimate is conservative regardless of whichever initial conditions are chosen  on the domain boundary $\partial \G_R$.

 Thus, the convergence time $\T_2$ of the closed-loop system \eqref{eq:SISO}--\eqref{eq:u2def}, initialized at  $O_4$, is bigger than the convergence time  $T_2$ of \eqref{eq:SISO}--\eqref{eq:u2def}, initialized on  \eqref{eq:BG1def}, i.e.,
 \begin{equation}
 \T_2\geq T_2,
 \label{comp}
 \end{equation}
 and it is given by \cite[relation (5.37)]{Orlov2012}
 \begin{equation}
	\label{eq:T2simplified}
	\T_2=\frac{\delta\left( \sqrt{1-\eta^2}+1\right) }{(\mu_1-N)\sqrt{1-\eta^2}}\; 
\end{equation}
where
 \begin{equation}
	\label{eq:tau1param}
		\eta=\frac{\mu_1-N}{\mu_2}< \frac{1}{\beta}. %\; \text{\cite[(relation (5.26)]{Orlov2012}}.
\end{equation}

Relations \eqref{eq:T2simplified} and \eqref{eq:tau1param}, coupled together, ensure \cite[cf. relation (5.40)]{Orlov2012} that
\begin{equation}
	\label{eq:ineqT2}
		 \T_2= \frac{\delta\left( \sqrt{1-\eta^2}+1\right) }{(\mu_1-N)\sqrt{1-\eta^2}}\leq  \frac{2\delta}{(\mu_1-N)\sqrt{1-\beta^{-2}}}. 
\end{equation}
Employing \eqref{comp} and taking into account that under condition
 \eqref{eq:C1}, imposed on $\mu_1$, one has
\begin{equation}
	\label{eq:upperBoundf}
			\frac{2\delta}{(\mu_1-N)\sqrt{1-\beta^{-2}}}\leq \T_s,
\end{equation}
  it follows that
  \begin{equation}
	\label{eq:upperBound}
		T_2\leq \T_2\leq \T_s.
\end{equation}
Hence, the convergence time  $T_2$ of the closed-loop system \eqref{eq:SISO}--\eqref{eq:u2def}, initialized on the domain boundary \eqref{eq:BG1def}, is smaller than the prescribed time instant $\T_s$. The proof of Theorem \ref{teo} is thus  completed.

\section{Supporting simulations}

Capabilities of the tuning of the twisting controller gains, resulting from Theorem \ref{teo},  are  illustrated in the numerical study of the fixed-time regulation problem of a simple pendulum, initialized in a prespecified domain. Consider a simple pendulum governed  by
\begin{equation} \label{eq:pend}
		\ddot q=b\left[ \tau(q,\dot q,t)-f_v \dot q-mgl\sin(q)+d(t) \right]
\end{equation}
where $b=1/(m_pl^2+J)$ and $q$ is the angular position. Hereinafter, $m_p$ is the mass of the pendulum, $l$ is the distance from its rotation axis to its center of mass, $J$ is the moment of inertia of the pendulum with respect to its center of mass, $f_v$ is the viscous friction coefficient, $g$ is the gravity acceleration, $\tau \in \RE$ is the torque produced by the actuator, the unknown term  $d(t)$ stands to account for external uniformly bounded disturbances.
%, restricted to the inequality given in Assumption~\ref{A:bound}.
The parameters of the pendulum  \eqref{eq:pend} are given in Table~\ref{t:planta}.
\begin{table}[htbp]
	\begin{center}
		\caption{Parametric values of the pendulum driven by actuator}
		\begin{tabular}{|l| |l| |l|}
			\hline
			parameter & value & unit  \\
			\hline \hline
			$m$ & 0.0474 & kg \\ \hline
			$l$ & 0.11 & m \\ \hline
			$J$ & $3.11\times 10^{-03}$ & kg m$^{2}$\\ \hline
			$g$ & 9.81 & m/s$^{2}$ \\ \hline
			$f_v$ & $2.43\times 10^{-04}$ & N s/rad \\ \hline
		\end{tabular}
		\label{t:planta}
	\end{center}
\end{table}
The pendulum model \eqref{eq:pend}, represented in terms of the state  errors $x_1(t)=q(t)-r,\ x_2(t)=\dot q(t)$ , takes the form
\begin{equation}\label{eqpende}
	\begin{split}
		\dot x_1&=x_2\\
		\dot x_2&=b \left[ \tau-f_v x_2-mgl\sin(x_1+r)\right]+\omega,
	\end{split}
\end{equation}
where $x=(x_1,x_2)^T\in \RE^2$, $r\in \RE$ is the desired position for the pendulum such that the initial error state vector $x^0=(x_1^0,x_2^0)\in \RE^2$ with $x_1^0=q(t_0)-r$ and $x_2^0=\dot q(t_0)$ satisfies $x^0\in\G_R$ and
\begin{equation}
	\omega(t)=bd(t).
	\label{varphipendulum}
\end{equation}
Setting
\begin{equation}
	\label{eq:tau}
	\tau=\frac{1}{b}u+f_v x_2 +mgl\sin(x_1+r),
\end{equation}
the control input \eqref{eq:tau} is then composed  of the friction-gravitation compensator $f_v x_2+mgl\sin(x_1+r)$ and the twisting algorithm $u(x_1,x_2)$ to be designed as in  \eqref{eq:u2def}.  Substituting  \eqref{eq:tau} in \eqref{eqpende} for $\tau$ yields
\begin{equation}\label{eq:CLu}
	\begin{split}
		\dot x_1&=x_2\\
		\dot x_2&=u(x_1,x_2)+\omega(t).
	\end{split}
\end{equation}
The  resulting system \eqref{eq:CLu} is consistent with the underlying system \eqref{eq:SISO} provided that  $\omega(t)$ is a uniformly bounded disturbance entering the system.

The control objective is that the closed--loop trajectories \eqref{eq:pend}, \eqref{eq:tau} reach the origin of the state space in a prescribed time $\T_s$ for any initial error $x^0\in\G_R$, regardless of whichever uniformly bounded disturbance $\omega(t)$ affects the system.

The simulations are carried out with several initial conditions for the pendulum as $q(0)=\{0,\frac{0.9R}{\mu_2}\}$ and $\dot q(0)=\{0.8\sqrt{2R},0\}$. With this  initial conditions we guaranties that $x^0\in\G_R$ in both cases. The desired value for the reference $r$ is set to zero, i.e., $r=0$. The unknown function $d(t)$ in \eqref{varphipendulum} is modeled as
\begin{equation}
	\label{eq:hit}
	d(t)=A\sin(w t)
\end{equation}
where $A=7\times 10^{-4}$  and $w=2\; rad/s$, thus the disturbance $\omega(t)$ given in \eqref{varphipendulum} is uniformly bounded as $\left|\omega(t) \right|\leq 0.19$.

\begin{figure}[H]
	\begin{center}
		\includegraphics[width=2.5in]{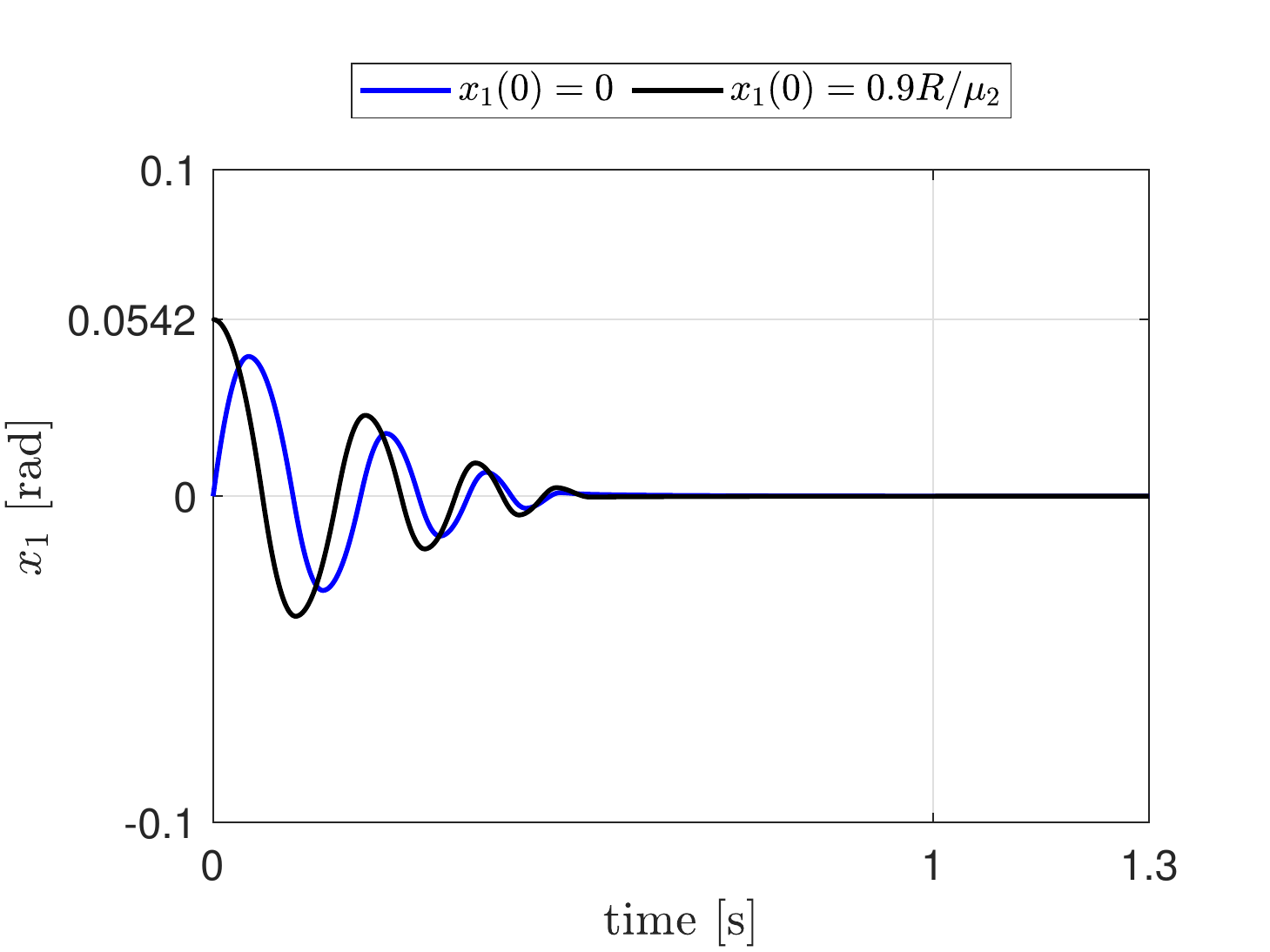}
		\includegraphics[width=2.5in]{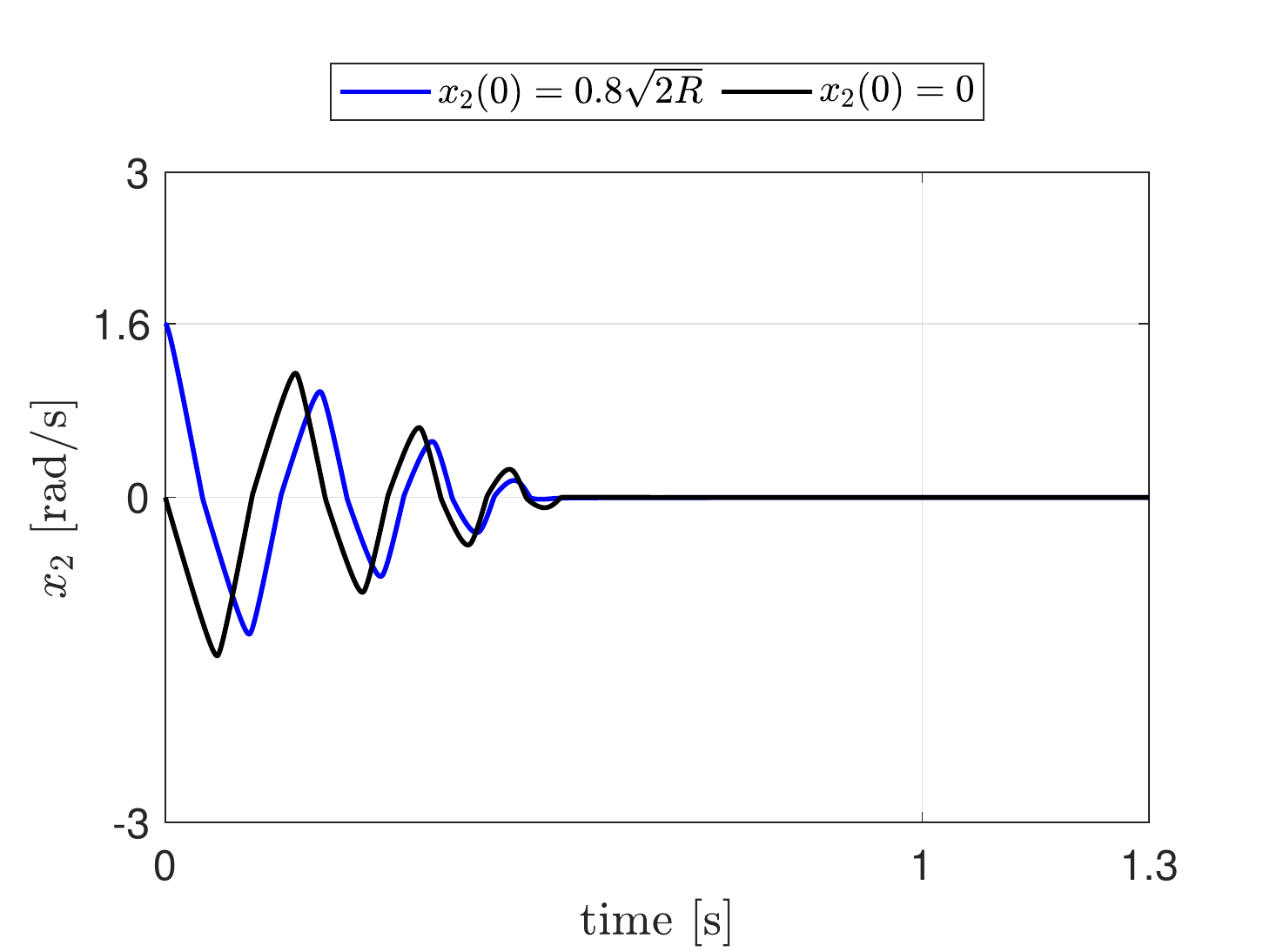}
		\includegraphics[width=2.5in]{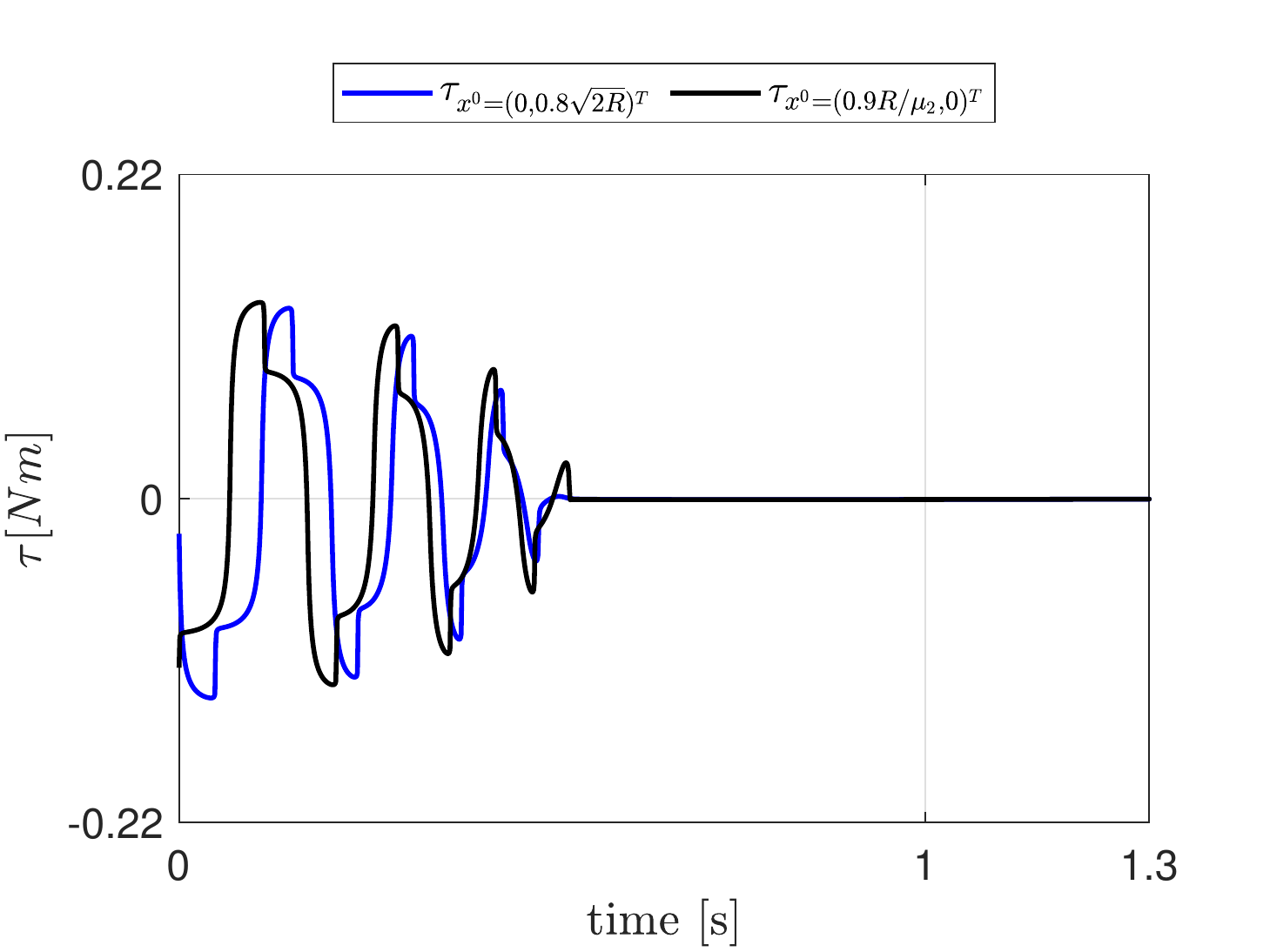}
	\end{center}
	\caption{Simulation results: time responses of the pendulum position error, velocity error, and control input.}
	\label{f:x1x2sim}
\end{figure}

\begin{figure}[H]
	\begin{center}
		\includegraphics[width=3.25in]{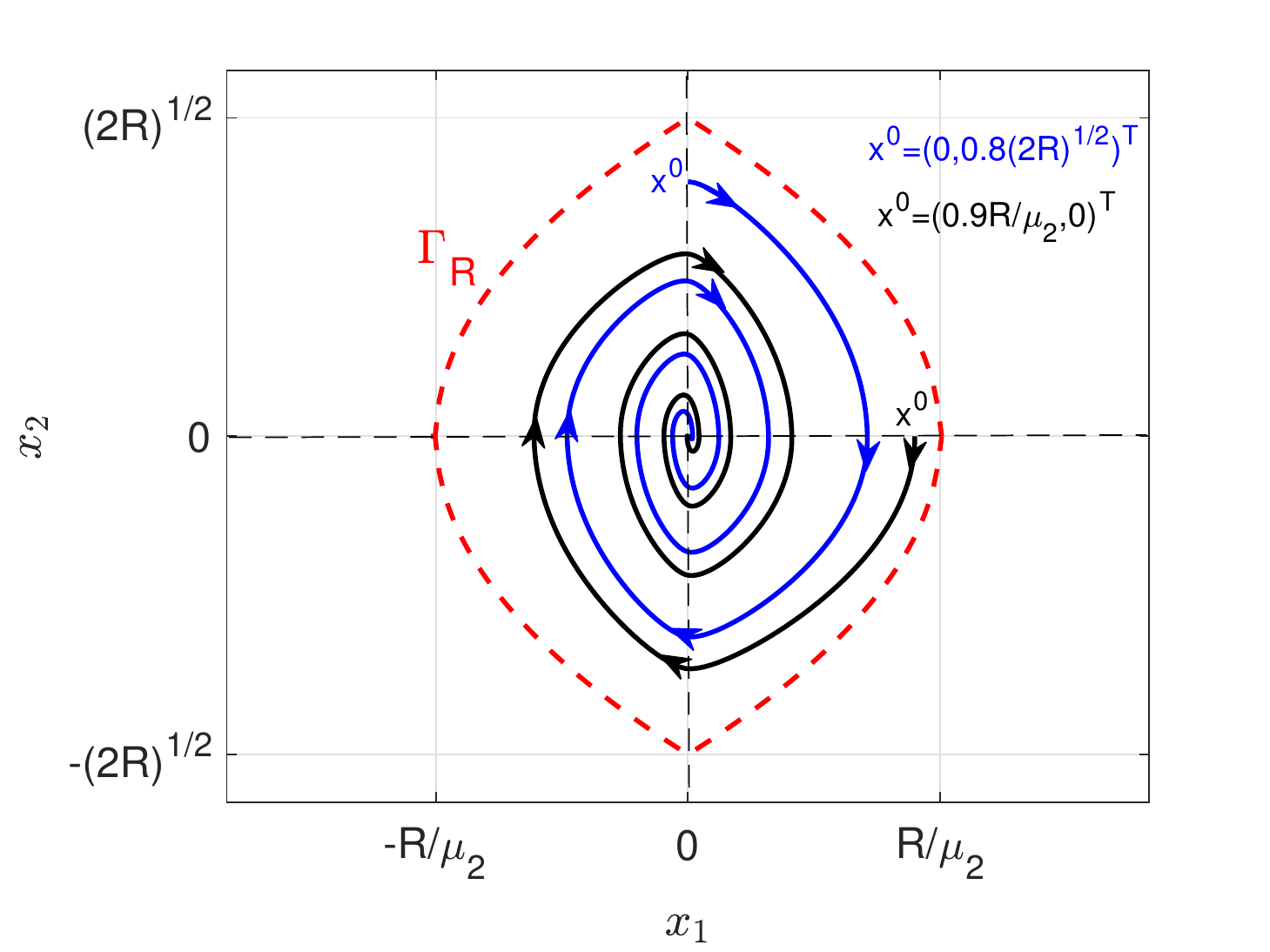}
	\end{center}
	\caption{Simulation results: a region near the origin of the phase-portrait illustrating the set $\G_R$.}
	\label{f:ppsim}
\end{figure}

The initial domain $\G_R$  in \eqref{eq:G1def} is specified with $R=2$, and the prescribed settling time is set to $\T_s=1$\,s. Taking $N=0.2$, and using the Theorem \ref{teo}, the tuning variables \eqref{eq:Rbeta} are selected as
\begin{equation}\label{eq:param}
	\begin{array}{ll}
	 R=2,& \beta=5,\\
    \rho=0.5,&\delta=3.1.
	\end{array}
\end{equation}
Then, using \eqref{eq:C1}--\eqref{eq:C2} together with \eqref{eq:param}, the twisting algorithm parameters are set to
\begin{equation}
	\label{eq:twistingGainSim}
	\mu_1=6.63, \qquad  \mu_2=33.24.
\end{equation}
As it can be seen in the Figure~\ref{f:x1x2sim}, the closed--loop system \eqref{eq:u2def}, \eqref{eq:pend}, \eqref{eq:tau} escape to the origin $(x_1,x_2)=0$ in a prescribed--time $t\leq \T_s$ irrespective of the initial conditions $x^0\in \G_R$ and  the uniformly bounded disturbances entering the system. It can additionally  be concluded  from the phase portrait of Figure~\ref{f:ppsim} that once in $\G_R$, the trajectories of the closed-loop system \eqref{eq:u2def}, \eqref{eq:pend}, \eqref{eq:tau}  never leave the level set $\G_R$ of the Lyapunov function  \eqref{eq:LyapTwisting}.

\section{Conclusions}
In the present note, a tuning procedure is formalized for the fixed time stabilization of the perturbed double integrator, initialized on an  {\it a priori} given domain and driven by the twisting algorithm.
The resulting tuning procedure is supported by numerical simulations enriching that of  \cite{Orlov2012}.
%%%%%%%%%%%%%%%%%%%%%%%%%%%%%%%%%%%%%%%%%%%%%%%%%%%%%%%%

%%%%%%%%%%%%%%%%%%%%%%%%%%%%%%%%%%%%%%%%%%%%%%%%%%%%%%%%%%%%%%%%%%%%%%%%%%%%%%%%%%%
\bibliography{BibOkAll}
\end{document}